\documentclass[twocolumn,twoside,slac_two]{revtex4}
\usepackage{graphicx}
\usepackage{fancyhdr}
\def\be{\begin{equation}}
\def\ee{\end{equation}}
\def\bea{\begin{eqnarray}}
\def\eea{\end{eqnarray}}
\def\nnb{\nonumber}
\newcommand{\scs}{\scriptscriptstyle}
\newcommand{\f}{\frac}

\begin{document}
\title{QCD Calculations of Radiative \boldmath{$B$} Decays}
\author{M. Misiak}
\affiliation{Institute of Theoretical Physics, Warsaw University, Warsaw, Poland}
\begin{abstract}
  The current status of $\bar{B} \to X_s \gamma$ decay rate calculations is summarized.
  Missing ingredients at the NNLO level are listed. The global normalization
  factor and non-perturbative effects are discussed. Arguments are presented
  that results for the cutoff-enhanced perturbative corrections have been
  misused in Ref.~\cite{Becher:2006pu} by applying them in the region
  $E_\gamma \in [1.0, 1.6]\,$GeV, which means that the corresponding numerical
  effect on ${\cal B}\left( \bar{B} \to X_s \gamma,~ E_{\gamma} > 1.6\,{\rm GeV}\right)$
  is unreliable.
\end{abstract}
\maketitle
\thispagestyle{fancy}
\section{Introduction \label{sec:intro}}

The motivation for precision studies of radiative $B$ decays is well known.
First, they are sensitive to new physics loop effects that often arise at the
same order in the electroweak couplings as the leading Standard Model (SM)
contributions.  Moreover, the inclusive $\bar{B} \to X_s \gamma$ rate is well
approximated by the perturbatively calculable radiative decay rate of the
b-quark. The CLEO~\cite{Chen:2001fj}, BELLE~\cite{Abe:2001hk} and
BABAR~\cite{Aubert:2005cu} measurements have been combined by
HFAG~\cite{Barberio:2008fa} to get
\be \label{bexp}
{\cal B}\!\left( \bar{B} \to X_s \gamma \right)_{\rm exp}
= ( 3.52 \pm 0.23 \pm 0.09 ) \times 10^{-4}
\ee
for $E_\gamma > 1.6\,$GeV. The corresponding SM prediction 
that was published two years ago\footnote{
More recent contributions are discussed in Secs.~\ref{sec:normal}--\ref{sec:nonpert}.} 
reads~\cite{Misiak:2006zs}
\be \label{bth}
{\cal B}\!\left( \bar{B} \to X_s \gamma \right)_{\rm SM}
= ( 3.15 \pm 0.23 ) \times 10^{-4}.
\ee
Its consistency with Eq.~(\ref{bexp}) provides strong constrains on 
many extensions of the SM (see, e.g., Ref.~\cite{Olive:2008vv}).

Resummation of large logarithms $\left(\alpha_s \ln M_W^2/m_b^2\right)^n$ in
the calculation of the decay rate is most conveniently performed after
decoupling the electroweak bosons and the top quark. In the resulting
effective theory, the relevant flavour-changing weak interactions are given by
a linear combination of dimension-five and -six operators\footnote{
  The specific matrices $\Gamma_i$ and $\Gamma'_i$ can be found in Ref.~\cite{Chetyrkin:1996vx}.}
\bea 
O_{1,2} &=& (\bar{s} \Gamma_i c)(\bar{c} \Gamma'_i b), \hspace{15.5mm}
\begin{array}{l} \mbox{\footnotesize (current-current} \\[-1mm] 
                 \mbox{\footnotesize ~operators)} \end{array}\nnb\\
O_{3,4,5,6} &=&  (\bar{s} \Gamma_i b) {\textstyle \sum_q} (\bar{q} \Gamma'_i q), \hspace{1cm}
\begin{array}{l} \mbox{\footnotesize (four-quark} \\[-1mm] 
                 \mbox{\footnotesize ~penguin operators)} \end{array}\nnb\\
O_7 &=& \f{e m_b}{16 \pi^2}\, \bar{s}_L \sigma^{\mu \nu} b_R F_{\mu \nu}, \hspace{6.5mm}
\begin{array}{l} \mbox{\footnotesize (photonic dipole} \\[-1mm] 
                 \mbox{\footnotesize ~operator)} \end{array}\nnb\\
O_8 &=& \f{g m_b}{16 \pi^2}\, \bar{s}_L \sigma^{\mu \nu} T^a b_R G^a_{\mu\nu}. \hspace{2mm}
\begin{array}{l} \mbox{\footnotesize (gluonic dipole} \\[-1mm] 
                 \mbox{\footnotesize ~operator)} \end{array} \label{ops}
\eea 
One begins with perturbatively calculating their \linebreak Wilson
coefficients $C_i$ at the renormalization scale $\mu_0 \sim (M_W, m_t)$. Next,
the Renormalization Group Equations (RGE) are used for the evolution of $C_i$
down to the scale $\mu_b \sim m_b/2$. Finally, the operator on-shell matrix
elements are calculated at $\mu_b$.\newpage

The Wilson coefficient RGE are governed by Anomalous Dimension Matrices
(ADM's) that are derived from ultraviolet divergences in the Feynman diagrams
with operator insertions. Around 20000 four-loop diagrams with $O_{1,2}$
insertions have been calculated in Ref.~\cite{Czakon:2006ss} to make the large
logarithm resummation complete up to~
${\cal O}\left[ \alpha_s^2 \left(\alpha_s \ln M_W^2/m_b^2\right)^n \right]$,
i.e. at the Next-to-Next-to-Leading-Order (NNLO) in QCD. Including such
corrections is necessary to suppress the theoretical uncertainty in
Eq.~(\ref{bth}) down to the level of the experimental error in
Eq.~(\ref{bexp}).  The numerical effect of the four-loop ADM's
on the branching ratio amounts to around $-4\%$ for $\mu_b = 2.5\,$GeV.  At
present, all the relevant Wilson coefficients $C_i(\mu_b)$ are known at the
NNLO~\cite{Czakon:2006ss,Bobeth:1999mk}.  However, evaluation of the matrix
elements at this order is still in progress --- see Sec.~\ref{sec:melem}.

\section{The global normalization factor \label{sec:normal}}

In order to reduce parametric uncertainties stemming from the CKM angles as
well as from the $c$- and $b$-quark masses, one writes the branching ratio as
follows \cite{Gambino:2001ew}
\bea \label{main}
{\cal B}\!\left(\bar{B} \to X_s \gamma\right)_{\scs E_{\gamma} > E_0}
&=& {\cal B}\!\left(\bar{B} \to X_c e \bar{\nu}\right)_{\rm exp} 
\left| \f{ V^*_{ts} V_{tb}}{V_{cb}} \right|^2 \times\nnb\\
&\times& \f{6 \alpha_{\rm em}}{\pi\;C} \left[ P(E_0) + N(E_0) \right],
\eea
where $\alpha_{\rm em} = \alpha_{\rm em}^{\rm on~shell}$, and
$N(E_0)$ denotes the non-perturbative correction (see Sec.~\ref{sec:nonpert}).
The $m_c$-dependence of $\bar{B} \to X_c e \bar\nu$ is accounted for by
\be \label{phase1}
C = \left| \f{V_{ub}}{V_{cb}} \right|^2 
\f{\Gamma\!\left(\bar{B} \to X_c e \bar{\nu}\right)}{
   \Gamma\!\left(\bar{B} \to X_u e \bar{\nu}\right)},
\ee
while $P(E_0)$ is defined by the perturbative ratio
\be \label{pert.ratio}
\f{\Gamma\!\left( b \to X_s \gamma\right)_{E_{\gamma} > E_0}}{
|V_{cb}/V_{ub}|^2 \; \Gamma\!\left( b \to X_u e \bar{\nu}\right)} = 
\left| \f{ V^*_{ts} V_{tb}}{V_{cb}} \right|^2 
\f{6 \alpha_{\rm em}}{\pi} \; P(E_0).
\ee
The NNLO expression for the phase-space factor\linebreak $C$ (\ref{phase1}) is
a known function of $m_c/m_b$ as well as non-perturbative parameters that
affect $N(E_0)$, too. All these quantities are determined in a single fit from
the measured spectrum of the inclusive semileptonic decay $\bar{B} \to X_c e
\bar{\nu}$.  The fits are performed using either the 1S or the kinetic
renormalization schemes. The corresponding results for $C$ and $m_c$ read
\be
\left( \begin{array}{c} C \\ m_c(m_c) \end{array} \right) =
\left\{ \begin{array}{ll}
\left( \begin{array}{l} 0.582 \pm 0.016 \\ 1.224 \pm 0.057 \end{array} \right), &
\mbox{ 1S \cite{Bauer:2004ve}},\\[3mm]
\left( \begin{array}{l} 0.546^{+0.023}_{-0.033} \\[1mm] 1.267 \end{array} \right), &
\mbox{ kin. \cite{Gambino:2008fj,Buchmuller:2005zv}}.\\[3mm]
\end{array} \right.
\ee
The above $\overline{\rm MS}$-scheme values of $m_c$ have been obtained from
the 1S- and kinetic-scheme ones using the three-loop and two-loop relations,
respectively. The three-loop relation for the kinetic scheme is not yet known.
The ratio $C$ is scheme-independent, but it is affected by the so-called weak
annihilation contribution $B_{\rm WA}$ that remains unknown. Since $B_{\rm
  WA}$ cancels out in Eq.~(\ref{main}), fixing its value is a matter of
convention in the present context. Here, we follow the convention of
Ref.~\cite{Gambino:2008fj}, namely $B_{\rm WA}(\mu = m_b/2)=0$.

The difference between the two determinations of $C$ amounts to $1.6\sigma$
when counted in terms of the upper error of the very recent kinetic-scheme
result \cite{Gambino:2008fj}. It is a consequence of using different
experimental data sets, methodology and renormalization schemes.  Fortunately,
the effects of changing $C$ and $m_c$ partially compensate each other in
Eq.~(\ref{main}) because $\partial/\partial m_c\; P(E_0) < 0$. For
$E_0=1.6\,$GeV, I find
\be \label{bbth}
{\cal B}\!\left( \bar{B} \to X_s \gamma \right)
= \left\{ \begin{array}{ll}
( 3.15 \pm 0.23 ) \times 10^{-4}, & \mbox{ 1S },\\[1mm]
( 3.25 \pm 0.24 ) \times 10^{-4}, & \mbox{ kin. },
\end{array} \right.
\ee
where the first result is just that of Ref.~\cite{Misiak:2006zs}, while the
second one has been obtained using the same code but with the input parameters
from Refs.~\cite{Gambino:2008fj,Buchmuller:2005zv}.

The actual value of ${\cal B}\left( \bar{B} \to X_s \gamma \right)$ in
Ref.~\cite{Gambino:2008fj} is somewhat larger than $3.25 \times 10^{-4}$
because $P(E_0)$ was calculated there using the one-loop rather than two-loop
determination of $m_c(m_c)$ from $m_c^{\rm kin}$. In principle, using the
one-loop relation is allowed at ${\cal O}(\alpha_s^2)$ because $P(E_0)$
becomes $m_c$-dependent only at ${\cal O}(\alpha_s)$.

\section{Cutoff-enhanced corrections \label{sec:cutoff}}

The perturbative ratio $P(E_0)$ in Eq.~(\ref{pert.ratio}) depends on the
cutoff energy $E_0$ via the dimensionless parameter
\be
\delta = 1 - \f{2 E_0}{m_b}.
\ee
For very small $\delta$, i.e. close to the kinematical endpoint $E_0=m_b/2$,
the usual (``fixed-order'') perturbative expansion breaks down because the
corrections behave like powers of $\ln\delta$. In that region, one needs to
resum large logarithms of $\delta$. Such a resummation of the cutoff-enhanced
corrections (i.e. corrections enhanced by powers of $\ln\delta$) has been
performed up to the NNLO in Refs.~\cite{Becher:2005pd,Becher:2006pu}. These
results constitute a valuable contribution to our knowledge of the photon
energy spectrum in the endpoint region.  However, they need to be treated with
extreme care further from the endpoint, where logarithms of $\delta$ no longer
dominate.  Naively, one might expect that resummation of small logarithms does
not hurt, even if it is not an improvement.  Unfortunately, this is not the
case because the logarithmic and non-logarithmic terms undergo a strong
cancellation away from the endpoint.
\begin{figure}[t]
\centering
\includegraphics[width=80mm]{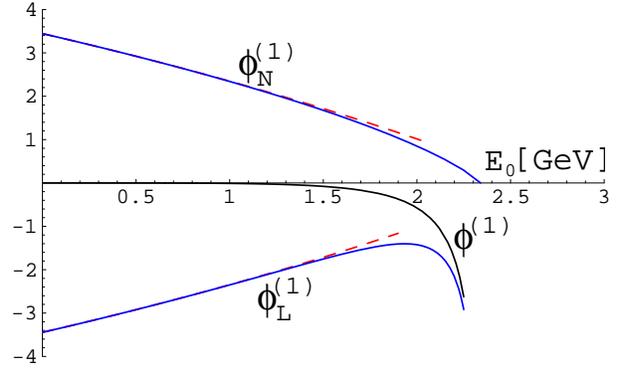}
\caption{Approximate cancellation of the logarithmic ($\phi^{(1)}_L$) and 
  non-logarithmic ($\phi^{(1)}_N$) terms in $\phi^{(1)}$ away from the
  endpoint.  The exact expressions from Eq.~(\ref{split}) are represented by
  the solid lines. The Taylor expansions of $\phi^{(1)}_L$ and $\phi^{(1)}_N$
  around ~$E_0=0$~ up to ~${\cal O}(E_0^3)$~ are shown by the dashed lines. \label{fig:phi1}}
\end{figure}

In order to illustrate this issue in a simple manner, let us consider only the
dominant photonic dipole operator $O_7$ in Eq.~(\ref{ops}). When all the other
operators are neglected, the fixed-order expression for the cutoff-dependence
of $P(E_0)$ is given by
\be \label{ppratio}
\f{P(E_0)}{P(0)} =
1 + \f{\alpha_s}{\pi  } \phi^{(1)}(\delta) +
    \f{\alpha_s^2}{\pi^2} \phi^{(2)}(\delta) + \ldots
\ee
Each of the functions $\phi^{(k)}$ can be split into two parts: $\phi^{(k)}_L$
that is polynomial in $\ln \delta$, and $\phi^{(k)}_N$ that vanishes at the
endpoint. The explicit expressions for $k=1$ read \cite{Ali:1990tj}
\bea
\phi^{(1)} &=& \phi^{(1)}_L + \phi^{(1)}_N,\nnb\\
\phi^{(1)}_L(\delta) &=& -\f{2}{3} \ln^2 \delta -\f{7}{3} \ln\delta - \f{31}{9},\nnb\\[1mm]
\phi^{(1)}_N(\delta) &=& \f{10}{3} \delta + \f{1}{3} \delta^2 - \f{2}{9} \delta^3 
                         + \f{1}{3} \delta (\delta-4) \ln\delta.~~~~~\label{split}
\eea
In Fig.~\ref{fig:phi1}, $\phi^{(1)}_L$, $\phi^{(1)}_N$ and their sum are
plotted as functions of $E_0$. The endpoint is located at $m_b/2 \simeq
2.34\,$GeV. One can see that $\phi^{(1)}_L$ begins to dominate around $E_0 =
2\,$GeV. On the other hand, already in the vicinity of $E_0 = 1.6\,$GeV, the
cancellation of the two components of $\phi^{(1)}$ is very strong.

\begin{figure}[t]
\centering
\includegraphics[width=80mm]{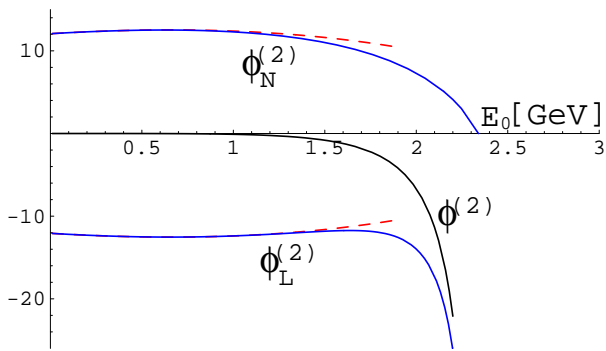}
\caption{Same as Fig.~\ref{fig:phi1} but for $\phi^{(2)}$ in the case when
  $\alpha_s = \alpha_s^{n_f=3}(m_b)$~ in Eq.~(\ref{ppratio}). \label{fig:phi2}}
\end{figure}

One may wonder whether similar cancellations occur at higher orders, too. A
positive answer concerning $\phi^{(2)}$ is immediate because this function can
easily be derived from the results of Ref.~\cite{Melnikov:2005bx}.  The
corresponding plot is presented in Fig.~\ref{fig:phi2} for the case when
$\alpha_s = \alpha_s^{n_f=3}(m_b)$~ in Eq.~(\ref{ppratio}). 

\begin{figure}[t]
\centering
\includegraphics[width=70mm]{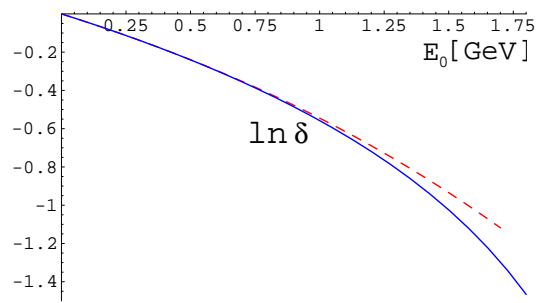}
\caption{$\ln \delta$ (solid) and its expansion (dashed) up to ~${\cal O}(E_0^3)$. \label{fig:log}}
\end{figure}

No explicit results at order ${\cal O}(\alpha_s^3)$ are available.  However,
we know on general grounds that all the $\phi^{(k)}$ behave like $\left( 2 E_0/m_b
\right)^4$ at small $E_0$.  Two powers of $E_0$ originate from $F_{\mu\nu}$ in
the vertex $O_7$ in Eq.~(\ref{ops}), and two additional powers come from the
phase-space measure $E_\gamma dE_\gamma$. Consequently, the Taylor expansions
of $\phi^{(k)}_L$ and $\phi^{(k)}_N$ at small $E_0$ up to ~${\cal O}(E_0^3)$~
must exactly cancel each other. These Taylor expansions are shown by the
dashed lines in Figs.~\ref{fig:phi1} and \ref{fig:phi2}.  One can see that
both $\phi^{(1)}_L$ and $\phi^{(2)}_L$ are well approximated by the dashed
lines in the region below $1.6\,$GeV. It must also be the case at higher
orders because $\phi^{(k)}_L$ are polynomial in ~$\ln\delta$~ that is well
approximated by the same expansion in the considered region (see Fig.~\ref{fig:log}).
Thus, cancellations like those shown in Figs.~\ref{fig:phi1} and
\ref{fig:phi2} are expected to occur at any order in $\alpha_s$
(see ``Note Added'').

In the approach of Refs.~\cite{Becher:2005pd,Becher:2006pu}, logarithms of
$\delta$ have been resummed at the NNLO in $\phi^{(k)}_L$, while
$\phi^{(k)}_N$ have been retained in the fixed order.
%
%
More precisely, Eq.~(\ref{ppratio}) has been effectively re-expressed as
\be \label{ppratioN}
\f{P(E_0)}{P(0)} =
X + \f{\alpha_s(\mu_b)  }{\pi  } \phi^{(1)}_N +
    \f{\alpha_s^2(\mu_b)}{\pi^2} \phi^{(2)}_N,
\ee
and $X$ has been calculated up to 
${\cal O}\left( \alpha_s^2\times \alpha_s^n \ln^m \delta  \right)$,
with~ $n=0,1,2,\ldots$,~ and~ $m=n,n+1,\ldots,m_{\rm max}(n)$.
This is a reasonable approximation only in the region very close to the
endpoint where no cancellation between the two components takes place.
Elsewhere, it leads to overestimating the numerical effect of the 
${\cal O}(\alpha_s^3)$ terms in Eq.~(\ref{ppratio}) by a factor of
order~
%
%
$\left| \phi^{(3)}_L/\phi^{(3)} \right| \sim \left( m_b/(2 E_0) \right)^4$ 
that amounts to around 4.6 for $E_0 = 1.6\,$GeV, and around 30 for $E_0 = 1.0\,$GeV.

Unfortunately, it was precisely the range\linebreak $E_\gamma \in [1.0, 1.6]\,$GeV where the
authors of Ref.~\cite{Becher:2006pu} applied their results to calculate the
effect on
${\cal B}\!\left( \bar{B} \to X_s \gamma,~ E_{\gamma} > 1.6\,{\rm GeV}\right)$.
They adopted the fixed-order result at $E_0 = 1.0\;$GeV from Ref.~\cite{Misiak:2006zs}
\be
{\cal B}\!\left( \bar{B} \to X_s \gamma,~ E_{\gamma} > 1.0\,{\rm GeV}\right) = 3.27 \times 10^{-4},
\ee
and supplemented it with their own numerical value of
$\left[ P(1.0)-P(1.6) \right]/P(0)$ 
that follows from Eq.~(\ref{ppratioN}). That value is almost twice
larger than in the fixed-order NNLO calculation.  In the end, their
result for the branching ratio with a cutoff at $E_0 = 1.6\,$GeV was
considerably lower than the one in Eq.~(\ref{bth}). In view of the
above remarks, their prediction should be considered unreliable.

\section{Status of the NNLO QCD calculations of the matrix elements \label{sec:melem}}

On the l.h.s.~of Eq.~(\ref{pert.ratio}) that defines $P(E_0)$, the\linebreak
denominator is already known at the NNLO~\cite{vanRitbergen:1999gs,Pak:2008qt}. 
In the expression for the numerator
\bea
\Gamma ( b \rightarrow X_s \gamma )_{E_{\gamma} > E_0} &=&
\frac{G_F^2 \alpha_{\rm em} m_b^5}{32 \pi^4} |V_{tb} V_{ts}^{\ast}|^2 
\times\nnb\\ &\times& 
\sum_{i,j=1}^8 C_i^{\rm eff} \, C_j^{\rm eff} \, G_{ij}(E_0),
\label{rate}
\eea
the quantities $G_{ij}$ are determined by the matrix elements of $O_1$,
\ldots, $O_8$.
%

\begin{figure}[t]
\centering
\includegraphics[width=23mm]{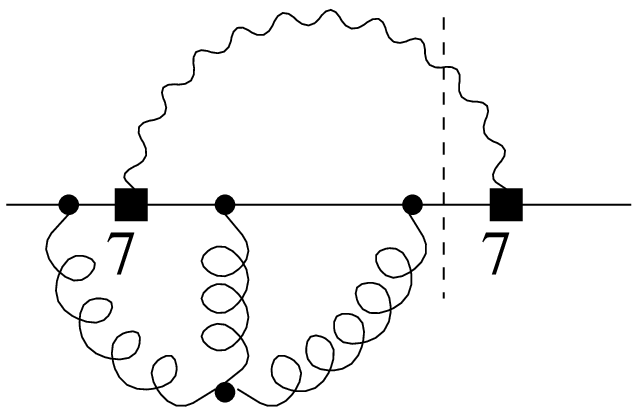}\hspace{13mm}
\includegraphics[width=23mm]{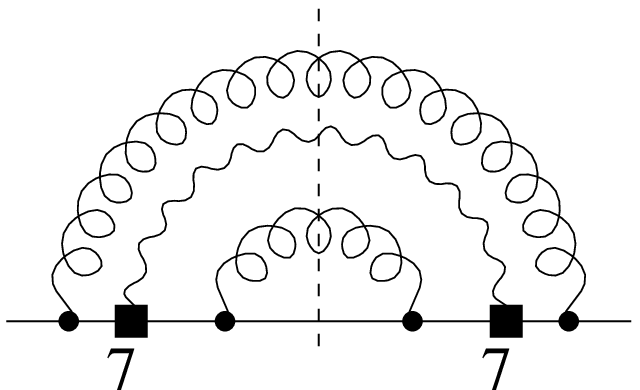}\\[2mm]
\includegraphics[width=35mm]{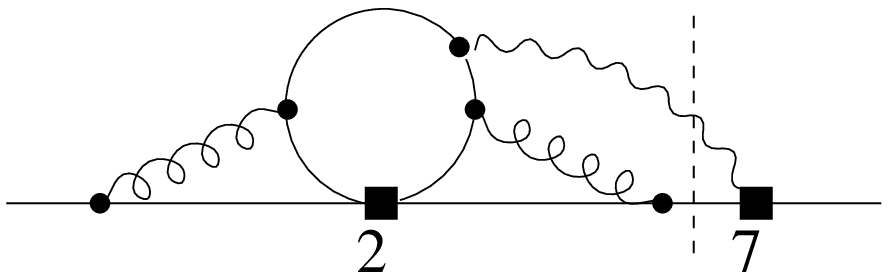}\hspace{5mm}
\includegraphics[width=25mm]{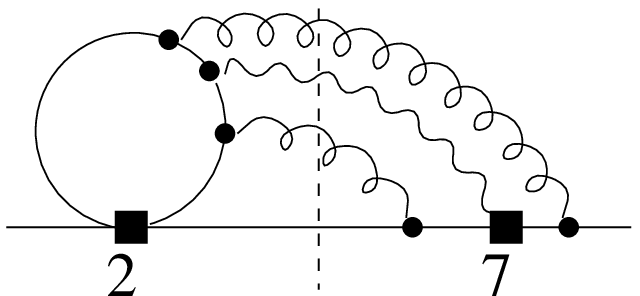}\\[2mm]
\includegraphics[width=35mm]{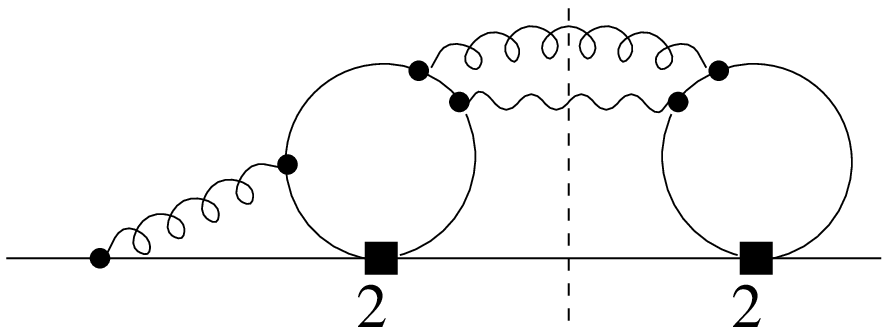}\hspace{5mm}
\includegraphics[width=25mm]{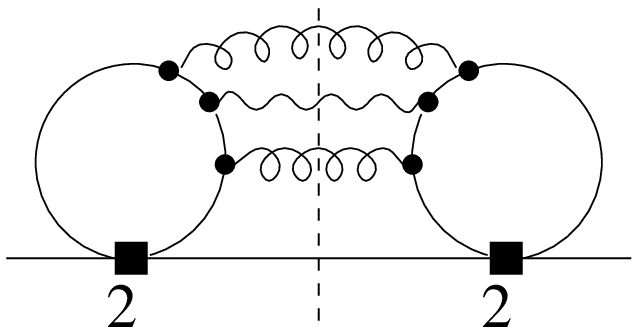}
\caption{Examples of Feynman diagrams that contribute to $G_{77}$, $G_{27}$ and
  $G_{22}$ at the NNLO. The dashed vertical lines mark the unitarity cuts.
  Black squares represent the operators $O_2$ and $O_7$ in Eq.~(\ref{ops}). \label{fig:Gij}}
\end{figure}

So far, only $G_{77}$ has been evaluated up to ${\cal O}(\alpha_s^2)$
in a complete manner~\cite{Melnikov:2005bx,Asatrian:2006ph,Asatrian:2006rq}.  
The remaining $G_{ij}$ are fully known at the Next-to-Leading Order
(NLO), i.e. up to ${\cal O}(\alpha_s)$ (see Ref.~\cite{Buras:2002er}
for a complete list of references). At the NNLO, it is practically
sufficient to restrict considerations to $G_{ij}$ with $i,j \in
\{1,2,7,8\}$ because the four-quark penguin operators have small
Wilson coefficients. It is often convenient to apply the optical
theorem, and calculate $G_{ij}$ by summing imaginary parts of the
$b$-quark propagator Feynman diagrams.  One can see in
Fig.~\ref{fig:Gij} that imaginary parts of three-, four- and five-loop
diagrams occur at the NNLO in $G_{77}$, $G_{27}$ and
$G_{22}$, respectively.

A relatively simple set of the NNLO contributions is given by diagrams with
quark loops on the gluon lines. The quark in the loop is either massive (charm
and bottom) or treated as massless (up, down and strange). Such contributions
are already
known~\cite{Asatrian:2006rq,Bieri:2003ue,Ligeti:1999ea,Boughezal:2007ny} for
all the $G_{ij}$ with $i,j \in \{1,2,7,8\}$, except for the massless case in
$G_{18}$ and $G_{28}$.  The BLM~\cite{Brodsky:1982gc} (or large-$\beta_0$)
approximation for the complete NNLO correction is derived from the massless
quark results~\cite{Bieri:2003ue,Ligeti:1999ea}.

In Ref.~\cite{Misiak:2006ab}, the asymptotic behaviour for $m_c \gg m_b/2$ was
calculated for all the non-BLM NNLO corrections to $G_{ij}$ with $i,j \in
\{1,2,7,8\}$, except for $G_{78}$ and $G_{88}$. Next, an interpolation in
$m_c$ of these corrections was performed assuming that the interpolated
quantities vanish at $m_c=0$.  The result of that procedure was an essential
input for the NNLO estimate in Eq.~(\ref{bth}). The overall error of around
7\% in the branching ratio was obtained by combining in quadrature four types
of uncertainties: 5\% non-perturbative, 3\% parametric, 3\% higher-order, and
3\% due to the $m_c$-interpolation ambiguity.

The results in Eqs.~(\ref{bth}) and (\ref{bbth}) do not include several
contributions to the branching ratio that are known at present. These
additional effects are summarized in Tab.~\ref{tab:add}. They sum up to around
$+1.6\%$, which is small when compared to the overall uncertainty of around
$7\%$. Therefore, Eq.~(\ref{bbth}) can still be treated as an up-to-date SM
prediction.
\begin{table}[t]
\begin{center}
\caption{Additional known effects not included in Eq.~(\ref{bbth}).}
\begin{tabular}{|l|l|} \hline
The BLM terms from Ref.~\cite{Ligeti:1999ea} & $+2.0\%$\\
$O_8$ in the 4-loop ADM's~\cite{Czakon:2006ss} & $-0.3\%$ \\
$b$ and $c$ loops on gluon 
lines~\cite{Pak:2008qt,Asatrian:2006rq,Boughezal:2007ny} & $+1.6\%$ \\
Non-perturbative ${\cal O}(\alpha_s \Lambda/m_b$) effects~\cite{Lee:2006wn} & $-1.5\%$ \\
Non-perturbative collinear effects~\cite{Kapustin:1995fk} & $-0.2\%$ \\\hline
Total & $+1.6\%$ \\\hline
\end{tabular}
\label{tab:add}
\end{center}
\end{table}

As the reader has already noticed, even in the $m_c$-interpolation approach of
Ref.~\cite{Misiak:2006ab}, there are still some missing NNLO ingredients,
namely:
\begin{itemize}
\item{} the BLM contributions to $G_{18}$ and $G_{28}$,
\item{} the large-$m_c$ results for $G_{78}$ and $G_{88}$.
\end{itemize}
Their numerical effect on the branching ratio is expected to remain within the
estimated higher-order uncertainty of around 3\%. The calculation of the most
interesting $G_{78}$ is very advanced~\cite{calc.78.yyy}, and the results
should become available soon (for any value of $m_c$). As far as $G_{88}$ is
concerned, its calculation in the large-$m_c$ limit will automatically give
the result for any value of $m_c$.

For the full NNLO calculation, the currently missing ingredients (apart from
the ones listed above) are the non-BLM corrections to
\begin{itemize}
\item[\it (i)] $G_{17}$ and $G_{27}$,
\item[\it (ii)] $G_{11}$, $G_{12}$ and $G_{22}$, 
\item[\it (iii)] $G_{18}$ and $G_{28}$.
\end{itemize}
The calculation of {\it (i)} in the $m_c=0$ limit is quite advanced, but also
extremely difficult and time-consuming --- see Ref.~\cite{Boughezal:2007km}
for the status reports. Once it is finished, the main challenge will amount to
finding {\it (ii)}, even for $m_c=0$.  The corrections {\it (iii)} are expected to
be numerically less important.

When the non-BLM corrections are known at\linebreak $m_c=0$ sometime in the future, the
interpolation in $m_c$ is still going to be necessary. However, our error
estimates should become more solid once the BLM approximation is no longer
used at the boundary. 

Finding all the non-BLM corrections for the actual value of $m_c \simeq m_b/4$
is even more difficult, but it must be considered at some point, too. A
calculation of the IR-divergent two-particle-cut contributions to {\it (i)} is
being currently performed for arbitrary $m_c$~\cite{Boughezal:2007km}.  The
IR-convergent two-particle-cut contributions to {\it (ii)} for arbitrary $m_c$
are already known because they are given by products of the NLO corrections.
The $(n\geq3)$-particle-cut contributions to {\it (ii)} vanish at the
endpoint, so their numerical relevance should be diminished by the high photon
energy cutoff.

Apart from the NNLO corrections, there are other perturbative contributions to
$\Gamma ( b \rightarrow X_s \gamma )_{E_{\gamma} > E_0}$ that have been
neglected so far, namely tree-level diagrams with the $u$-quark analogues of
$O_{1,2}$ and the four-quark penguin operators $O_{3,4,5,6}$. Such
contributions are suppressed with respect to the leading term by $|(C_{1,2}^u,
C_{3,4,5,6})/C_7|^2 \leq 0.2$, where $C_{1,2}^u = (V_{us}^* V_{ub})/(V_{ts}^*
V_{tb}) C_{1,2}$, as well as by the high photon energy cutoff.  A quantitative
verification of how small they really are should become available
soon~\cite{Kaminski:2008yyy}.

\section{Non-perturbative effects \label{sec:nonpert}}

Let us begin with considering a simplified world where $m_c=m_b$. There, in
the decay of the $\bar{B}$ meson, a high-energy photon ($E_\gamma \sim m_b/2$)
can be produced in four different ways:
\begin{itemize}
\item[1.] Hard: The photon is emitted directly from the hard process of the $b$-quark decay.
\item[2.] Conversion: The $b$-quark decays in a hard way into quarks and gluons only.
  Next, one of the decay products scatters in a non-soft radiative manner with the
  remnants of the $\bar B$ meson. This can be viewed as a parton-to-photon
  conversion in the QCD medium.
\item[3.] Collinear: In the process of hadronization, a collinear photon is emitted. 
\item[4.] Annihilation: An energetic $q\bar q$ state produced in the
  $\bar{B}$ meson decay disintegrates radiatively.
\end{itemize}

The hard way would be the only one if no other operators but $O_7$ were
present in the effective theory. Non-perturbative effects in such a case were
first analyzed in Ref.~\cite{Falk:1993dh}. They arise as corrections of
order $\Lambda^2/m_b^2$ to
$\Gamma ( b \rightarrow X_s \gamma )_{E_{\gamma} > E_0}$
when $(m_b-2E_0) \sim m_b$. Moreover, these ${\cal O}(\Lambda^2/m_b^2)$ terms
cancel out in $N(E_0)$ in Eq.~(\ref{main}) with the analogous non-perturbative
corrections to the charmless semileptonic rate. Thus, we are left with the
small ${\cal O}(\Lambda^3/m_b^3)$ effects \cite{Bauer:1997fe}. The 
${\cal O}\left(\Lambda^2/(m_b-2E_0)^2\right)$ corrections are also small
\cite{Neubert:2004dd} for $E_0 \leq 1.6\,$GeV. All these terms are included in
Eq.~(\ref{bbth}), and affect the branching ratio by around $-0.7\%$.

The photon production via conversion is suppressed both by $\alpha_s$ (due to
the non-soft scattering) and by $\Lambda/m_b$ (due to dilution of the target).
The analysis in Ref.~\cite{Lee:2006wn} shows that no other suppression factors
occur. An effect on the branching ratio of roughly $-1.5 \pm 1.5\%$ was found
in that paper (see Tab.~\ref{tab:add}).

In our simplified case ($m_c=m_b$), the tree-level decay 
$b \to ({\rm quarks\;\&\;gluons})_s$ is possible only via the operators
$O_{3,4,5,6,8}$, and by the $u$-quark analogues of $O_{1,2}$. Consequently,
the collinear photon emission is suppressed either by $\alpha_s |C_8/C_7|^2
\leq 0.08$, or by $|(C_{1,2}^u, C_{3,4,5,6})/C_7|^2 \leq 0.2$.  Moreover,
there is an additional suppression by products of the quark electric charges
and, most importantly, by the high photon energy cutoff. The results of
Ref.~\cite{Kapustin:1995fk} lead to an estimate that the non-perturbative
collinear effects due to $O_8$ amount to around $-0.2\%$ in the branching
ratio for $E_0 = 1.6\,$GeV (see Tab.~\ref{tab:add}).

As far as annihilation is concerned, photons originating from decays of
$\pi^0$, $\eta$, $\eta'$ and $\omega$ are removed on the experimental side as
(huge) background. 
%
%
Contributions from the other established $q\bar q$ mesons are negligible. 
The corresponding perturbative diagrams are responsible for only around 0.1\%
of the total rate for $E_0 = 1.6\,$GeV.

Once the assumption $m_c\!=\!m_b$ is relaxed, the ${\cal O}(\Lambda^2/m_b^2)$
hard effects from $O_{1,2}$ get replaced by a series of the form~\cite{Buchalla:1997ky}
\be
\f{\Lambda^2}{m_c^2} \sum_{n=0}^{\infty} b_n \left( \f{\Lambda m_b}{m_c^2} \right)^n,
\ee
with quickly decreasing coefficients $b_n$. The calculable leading term has
been included in Eq.~(\ref{bbth}). It affects the branching ratio by around
$+3.1\%$.
%
%
%

All the quantitatively estimated non-perturbative effects that have been
mentioned so far sum up to
\be
(-0.7-0.2-1.5+3.1)\% = +0.7\%.
\ee
However, since their evaluation is often very uncertain, and the knowledge of
${\cal O}(\alpha_s \Lambda/m_b)$ contributions is by no means complete, a
non-perturbative error of $\pm 5\%$ has been assumed in Eq.~(\ref{bbth}), as
already mentioned Sec.~\ref{sec:melem}. Probably the most interesting of all
the unknown ${\cal O}(\alpha_s \Lambda/m_b)$ effects originate from
charm annihilation in the massive $(\bar cs)(\bar qc)$ intermediate states
($q=u$ or $d$), for the actual value of $m_c \simeq m_b/4$. 
%

One should remember that the error in Eq.~(\ref{bexp}) is affected by
a non-perturbative theoretical uncertainty, too.  It follows from the fact that
the actual measurements are not performed with $E_0 \simeq 1.6\,$GeV. The most
precise experimental results correspond to significantly higher photon energy
cutoffs for which the
${\cal O}\left(\Lambda^n/(m_b-2E_0)^n\right)$ 
effects are no longer small.  These effects are described by a
non-perturbative shape function~\cite{Neubert:1993ch} that is constrained by
the semileptonic data.  The very recent analysis~\cite{Ligeti:2008ac} of this
function and its effects on the $\bar{B} \to X_s \gamma$ photon spectrum can
hopefully provide input for the future HFAG averages.

\section{Conclusions \label{sec:concl}}

Thanks to the efforts of the past years, the uncertainties in ${\cal
  B}\!\left( \bar{B} \to X_s \gamma \right)$ have reached the level of around
$\pm 7\%$ on both the experimental and theoretical sides. A significant
progress in the perturbative calculations is expected in the near future.
However, understanding the ${\cal O}(\alpha_s \Lambda/m_b)$ non-perturbative
effects remains the key issue.

\section{Note Added}

After the first version of this article was submitted to the arXiv, Einan
Gardi pointed out to me that the approximate cancellation of logarithmic and
non-logarithmic terms (Sec.~\ref{sec:cutoff})
has already been discussed in Ref.~\cite{Andersen:2006hr}.

\begin{acknowledgments}
I would like to thank the organizers of the\linebreak HQL~2008 conference for hospitality. 
I am grateful to P.~Gambino, C.~Greub and T.~Ewerth for helpful remarks.  
This work has been supported in part by the Polish Ministry of Science and
Higher Education as a research project N~N202~006334 (in years 2008-11), 
and by the EU-RTN Programme, contract no.~MRTN--CT-2006-035482, ``Flavianet''.
\end{acknowledgments}

\end{document}